# Role of strain on the stability of B, C, N, and O in Iron

Kishor Peddauvvala[1], Prince Gollapalli[1], Debolina Misra[3], Prajeet Oza[1],

Satyesh Kumar Yadav[1,2]

1 Department of Metallurgical and Materials Engineering, Indian Institute of Technology

(IIT) Madras, Chennai 600036, India.

2 Center for Atomistic Modelling and Materials Design, Indian Institute of Technology

(IIT) Madras, Chennai 600036, India.

3 Department of Physics, Indian Institute of Information Technology Design and Manufacturing (IITDM) Kancheepuram.

**Abstract:**

The preference for the occupation of solute atoms like B, C, N, and O at various sites in iron is generally explained by the size of the solute and the volume available for the solute atoms to occupy. Such an explanation based on the size of solute atoms and available space at the occupation site assumes that distortion alone dictates the stability of solute atoms. Using first-principles density functional theory (DFT), we separately calculate the distortion energy (DE) and electronic binding energy (EBE) of solute atoms in iron. We show that electronic binding dictates the relative stability of O rather than distortion. In contrast, the relative stability of B, C, and N is dictated by the distortion it exerts on iron atoms. Contribution to the relative stability of B atoms is dictated mostly by distortion. It suggests that B could occupy a large volume region like grain boundaries. The same agrees with experiments indicating B segregates at grain boundaries and planar defects. Such conclusions could not have been drawn from the formation energy calculation, which shows that B is stable at the substitution site.

**Introduction:**

Adding B, C, N, and O in ppm level can drastically change the properties hence the structure of Fe. For example, ppm level boron addition to micro-alloyed steels affects the hot flow behaviour through the gain refinement [1]. A large amount of C at interstitial sites increases elastic constants and bulk modulus. When it segregates at the dislocation core, it increases the Peierls barrier to the movement of dislocation. [2–6]

Occupation of B, C, N, and O atoms at various sites in Fe have been widely studied so that properties and structural transformation could be explained. Several thumb rules have been developed to explain the occupation of these solutes in Fe. Most of them assume the size available for occupation, and the size of the solute plays a vital role in deciding the site solute atoms occupy. B having the largest size is suggested to occupy the substitutional site as it provides large space. But several experimental observations indicate that B sits at grain boundaries. [7]

Formation energy calculated using DFT has been used to assess the relative stability of various solute atoms in Fe. Total formation energy cannot be used to comment on the role of strain exerted by interstitial solute atoms on their stability. To understand the role played by strain and electronic bonding, we calculate distortion energy (DE) and electronic binding energy (EBE) in interstitial solute atoms in iron. DE indicates the supercell's energy increase due to distortion of Fe atoms after introducing an interstitial solute atom.

**Results and discussion**

Interstitial formation energy was calculated using the following formula.

$$E^f(SA)_{Fe} = E(Fe_N+SA) - NE(Fe) - E(SA)$$

Where $E(Fe_N+SA)$ is the total energy of the Fe in B.C.C and F.C.C. crystal structures with interstitial either octa or tetra sites, $E(Fe)$ is the energy associated with the single Fe atom in defect-free B.C.C or F.C.C supercell, and $E(SA)$ energy of the single interstitial placed at the center of the 10x10x10 A° supercell.

VASP code[8] with the projector augmented [9] was used to implement the DFT. Generalized gradient approximations in the Perdew-burke-Ernzerhof [10] form was used as the exchange-correlation functional. To sample the K-points in the Brillouin zone, the Monkhorst -Pack scheme was used. A 4×4×4 and 3×3×3 K-point grids were used for the B.C.C and F.C.C supercells, respectively. An energy cut-off of 520 eV was used for all the calculations. All primitive cells were volume relaxed. However, to get the atomic positional change around solute atom site supercell volume was fixed.

Table 1 lists the solutes' interstitial formation energy at Octa, tetra, and substitutional sites in BCC and FCC iron. Formation energies agree with previously calculated values at the same level of theory [2]. Boron stabilizes at substitution, and C, N, and O stabilize at the octahedral site in both B.C.C and F.C.C iron.

**Table 1: Formation enthalpies, electronic binding energies, Distortion energies, Bader charges, volume and volume changes of sites. For single B, C, N, O, placed at two interstitial sites and at the B.C.C and F.C.C Iron supercell substitutional sites. Initial volumes for octahedral, tetrahedral, and substitution are 7.58, 1.9, and 22.73 respectively for B.C.C Iron and 6.83, 1.71, and 34.13 for F.C.C Iron.**

|  | Site | BCC-Fe | | | | FCC-Fe | | | |
|---|---|---|---|---|---|---|---|---|---|
|  |  | B | C | N | O | B | C | N | O |
| Formation-enthalpy(eV) | O | -5.53 | -7.17 | -5.05 | -3.73 | -6.25 | -7.86 | -5.51 | -3.82 |
|  | T | -4.85 | -6.32 | -4.35 | -3.25 | -3.55 | -5.36 | -4.11 | -3.77 |
|  | S | -5.64 | -4.80 | -2.04 | -2.16 | -6.53 | -5.48 | -2.26 | -1.86 |
| Bader charge | O | 0.88 | 0.96 | 0.87 | 1.18 | 0.56 | 0.95 | 1.14 | 1.03 |
|  | T | 0.75 | 0.89 | 1.03 | 1.08 | 0.79 | 0.98 | 1.2 | 1.14 |
|  | S | 0.6 | 0.98 | 1.12 | 1.1 | 0.68 | 0.8 | 1.03 | 1.07 |
| Volume(Å$^3$) | O | 9.73 | 9.14 | 9.04 | 9.68 | 9.48 | 8.63 | 8.48 | 8.8 |
|  | T | 2.89 | 2.62 | 2.53 | 2.7 | 3.28 | 2.87 | 2.76 | 2.89 |
|  | S | 20.61 | 19.75 | 20.03 | 20.97 | 32.98 | 32.85 | 34.55 | 35.99 |
| %Change in volume | O | 28 | 21 | 19 | 28 | 39 | 26 | 24 | 29 |
|  | T | 53 | 38 | 34 | 43 | 92 | 68 | 61 | 69 |
|  | S | -9.37 | -13.11 | -11.92 | -7.78 | -3 | -4 | 1 | 6 |
| Calculate atomic radii (Å) |  | 0.87 | 0.67 | 0.56 | 0.49 |  |  |  |  |

Distortion binding energy was calculated using the following equation.

$$E^d(I)_{Fe} = E(Fe_N\text{-}I) - NE(Fe), \quad E^d(S)_{Fe} = E(Fe_N\text{-}S) - E(Fe_{N-1})$$

$E(Fe_N\text{-}I)$ and $E(Fe_N\text{-}S)$ are the energy iron supercell having the same arrangement of iron atoms as obtained from the relaxed structure of supercell without a solute atom at interstitial and substitution positions, respectively. And $E(Fe_{N-1})$ energy iron supercell has the vacancy in it. $E^d(I)_{Fe}$, and $E^d(S)_{Fe}$ are distortion binding energies at the interstitial and substitution positions, respectively.

EBE indicates the bond strength of solute and Fe atoms if the distortion does not happen. It captures the strength of the bond at various sites.

Electronic binding energy was calculated using the following formula

$$E^e(I)_{Fe} = E^f(SA)_{Fe} - E^d(I)_{Fe}, \quad E^e(S)_{Fe} = E^f(SA)_{Fe} - E^d(S)_{Fe}$$

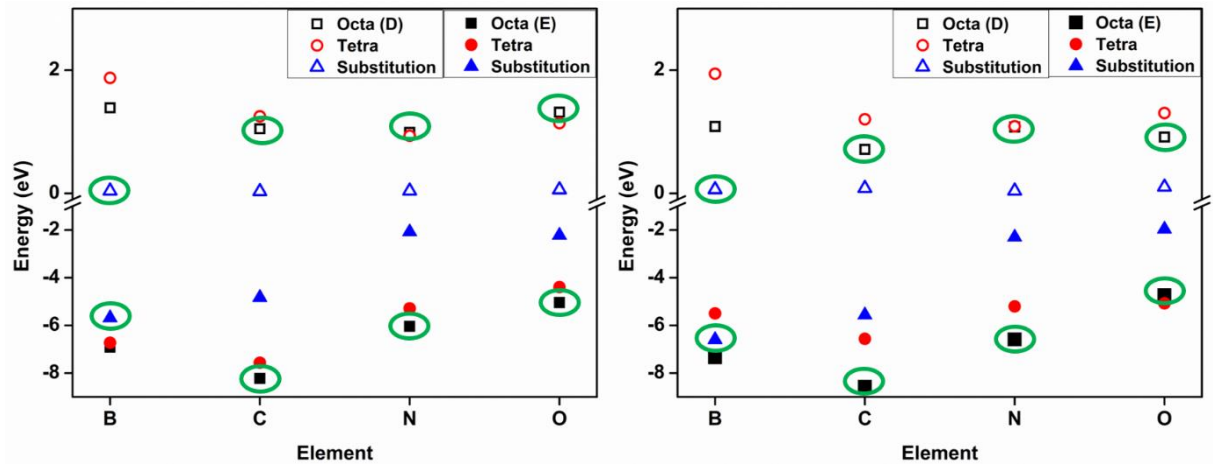

**Figure 1** Distortion energy (Open symbols) and electronic binding energy (filled symbols) associated with solute atoms when placed at octahedral, tetrahedral voids, and substitution position in a) B.C.C b) F.C.C iron. Solute atoms arranged from B to O are in the decreasing order of their calculated radius.

Figure 1 shows the DE and EBE associated with the different solute atoms in different sites. DE and EBE of the most stable site for the B, C, N, and O are enclosed with the green circles. It is widely accepted that the stable site for the C, N, and O is octahedral. With an explanation that the octahedral site has a large volume which gives less distortion assuming distortion to be a significant factor in dictating the stability of solute atoms [11,12]. But, figure 1 shows that DE is least at the substitution site, but C, N, and O are stable at the octahedral site. And also, their EBE is lowest at the octahedral site. Which tells that distortion is not only the dictating factor. On the other hand, Boron stabilized at the substitution despite the EBE not being the lowest. It indicates that boron stability in B.C.C and FCC iron depends on DE than EBE.

We find N and O occupies the octahedral site where distortion is higher. Though the calculated radius of the O is the lowest of all [13], O's octahedral site distortion binding energy is almost equal to that of B. Considering the hard-sphere model, distortion energy should decrease as the size of the atom decreases. Still, on the contrary, it shows that the oxygen atom creating distortion is equal to the boron, as boron consider to be the biggest among all interstitials. As an octahedral void is bigger than a tetrahedral void, the distortion produced by the interstitial in a tetrahedral void should be higher. The same was observed in the case of boron and carbon. In the case of nitrogen and oxygen, this has been reversed, as seen in the figure.

The volume enclosed by the center of the neighbouring atoms around the solute atom is considered the volume of the particular site. In the case of the octahedral site, six atoms surround the solute atom, four and that of the tetrahedral position, and eight in the case of the substitution site. Volume change has been calculated by taking the difference between the volumes of the site before and after the introduction of the solute atom. Volume contraction is indicated with a negative sign. DE at the substitutional site is close to zero in all cases. Volume should shrink around solute in the substitution site. But bonding is weaker at the substitution site comparatively. Hence there is no significant distortion/ or volume change compare to other sites.

It is expected that Boron is causing the maximum volume change among the other solute atoms as it is considered the biggest among all solute atoms. However, oxygen volume changes are almost equal to the boron in the case of both B.C.C and F.C.C octahedral sites. This is counter-intuitive as the calculated radius of the oxygen is the least. It causes the volume change equal to boron. To explain this, we carried Bader charge analysis on the solute atoms to know the effect of charge around the solute atom on its stability in the respective site.

Bader charge analysis was carried out on the solute atom[14]. It is a reliable method to estimate the charge on an atom. This method estimates the charge on an atom by considering the space around the atom as surfaces that run through the minima in the charge density, representing the Bader volume[15,16]. The difference between the valance electrons used for the DFT calculations to the calculated charge gives the charge on a solute atom.

Bader charge accumulation increases from boron to oxygen. In B.C.C, iron, where the charge accumulated oxygen is maximum, could be due to the partial ionization of atoms as the electronegativity increases from boron to oxygen[17]. Due to the accumulated charge around the oxygen, the effective radius of the oxygen has been changed, causing volume change equal to boron.

EBE gives a quantitative estimate of a solute's bond strength of interstitials at various sites. Neither FE nor EBE can be used to compare the relative bond strength of various solutes, as the energy of a single atom of solute dictates FE and EBE. For example, the EBE of the B atom is consistently lower at all sites compared to O, although it is well established that oxides' bond strength is much higher than borides.

C strongly depends on the environment, but if the environment remains the same, it would prefer to occupy a site with a lower strain. That is why formation energy of C in FCC is lower than BCC. For BCC Fe, it is interesting to note that the maximum difference in EBE of B at various sites is 1.24 eV, while the maximum difference in EBE for O, N, and C is 2.82 3.96 eV, and 3.39 eV, respectively. The environment dependence of EBE is driven by the strength of the bond and directional dependence of bonding. N has a maximum difference in EBE as it is the strongest covalent bond among B, C, and N. Drop is the maximum difference for O despite having a strong bond due to its ionic nature, which does not have strong direction dependence. A similar trend is seen in FCC iron as well. This indicates a very weak environment dependence of EBE for B compared to O, N, or C, which is possibly due to weak bonding as confirmed by DOS.

To elucidate further the bonding nature of the solute atoms and the iron, partial density of states (PDOS) was calculated. PDOS plotted between the solute atom and the average of the surrounding Fe atoms forming bonds with the solute atom in the respective site. Figure 2 shows hybridization from boron to oxygen is increasing. The figure shows that the boron P-orbital is not showing any hybridization with the iron d-orbital, which indicates the non-binding nature of the boron.

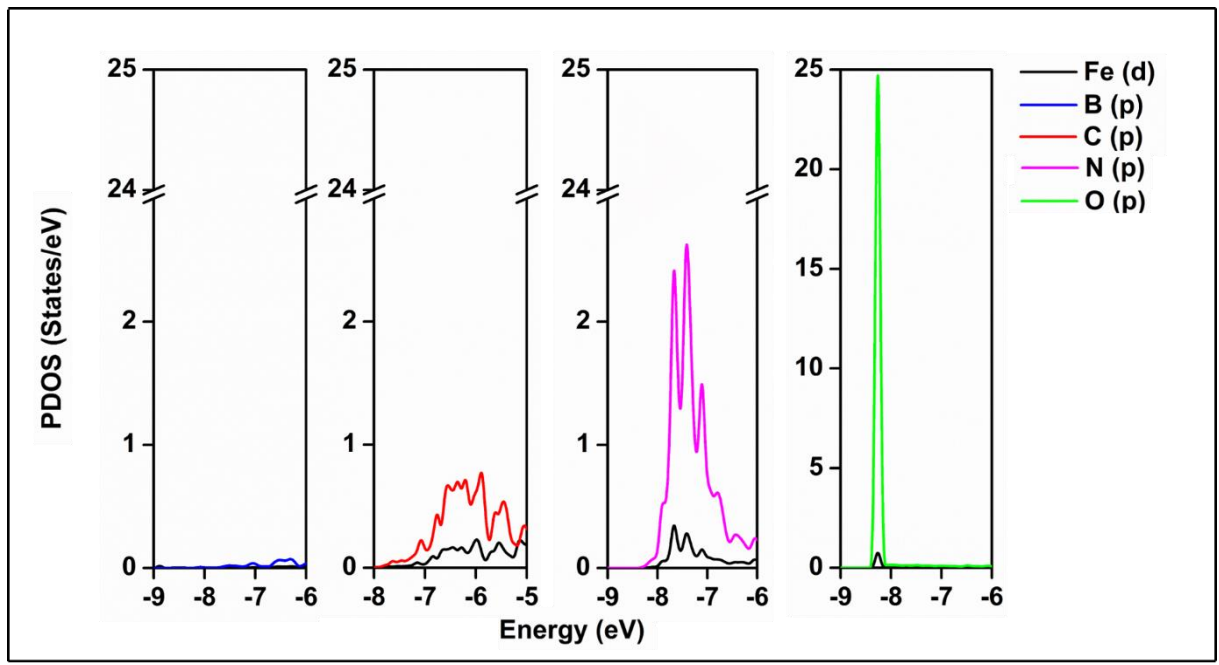

[3–6,18,19]

**Figure 2 partial density of states of solute atoms with the B.C.C iron representing the hybridization zone**

Weak dependence on the environment and the fact that B solute stability is strongly dictated by strain suggests that B will occupy a site with a large volume. Besides substitution sites with large volumes, B prefers to occupy grain boundaries or planar defects as they have a large volume. This is in line with several experimental and theoretical observations that B tends to segregate at grain boundaries [3–6,18,19]

**Conclusions:**

By separately calculating electronic binding energy and distortion energy of solute atoms in Fe using DFT, we show that strain alone does not play a role in the preference of a site of solute atoms. Results help us understand the limitations of the thumb rule (larger atoms will occupy a larger site and smaller ones will occupy a smaller site) on the solute atoms and the size of space available at various sites in Fe. While B follows the thumb rule, the genesis of O occupation of the octahedral site lies in the strong bonding to Fe in the octahedral environment. We show that B does not bind strongly to Fe and its stability is dictated by stain alone, which explains the preference of B to segregate at grain boundaries (large open volume hence less strain exerted). C, N, and O strongly prefer an octahedral environment, which helps us explain why C at the dislocation core rearranges the Fe atom around it so that it from a bond with 6 Fe atoms.